\documentclass[twoside,twocolumn,9pt]{article}
\usepackage{lmodern}
\usepackage{extsizes}
\usepackage[super,sort&compress,comma]{natbib} 
\usepackage[version=3]{mhchem}
\usepackage[left=1.5cm, right=1.5cm, top=1.785cm, bottom=2.0cm]{geometry}
\usepackage{balance}
\usepackage{times,mathptmx}
\usepackage{sectsty}
\usepackage{graphicx} 
\usepackage{lastpage}
\usepackage[format=plain,justification=justified,singlelinecheck=false,font={stretch=1.125,small,sf},labelfont=bf,labelsep=space]{caption}
\usepackage{float}
\usepackage{fancyhdr}
\usepackage{fnpos}
\usepackage[english]{babel}
\usepackage{array}
\usepackage{droidsans}
\usepackage{charter}
\usepackage[T1]{fontenc}
\usepackage[usenames,dvipsnames]{xcolor}
\usepackage{setspace}
\usepackage[compact]{titlesec}
\usepackage{upgreek}
\usepackage{hyperref}
\hypersetup{
    colorlinks=true,
    citecolor=blue,
    filecolor=black,
    linkcolor=blue,
    urlcolor=blue
}
\usepackage[capitalise,nameinlink]{cleveref}
\usepackage{overpic}
\usepackage{soul}
\usepackage{epstopdf}
\usepackage{siunitx}

\usepackage{nameref}
\usepackage[bottom]{footmisc}
\DeclareRobustCommand{\citepp}[1]{\citeauthor{#1}~\cite{#1}}
\usepackage{soul,color}

 \soulregister \cite7
 \soulregister \citep7
 \soulregister\ref7
 \soulregister\pageref7

\definecolor{cream}{RGB}{222,217,201}
\addto{\captionsenglish}{\renewcommand{\refname}{Notes and references}}

\crefname{equation}{Eqn}{Eqns}

\begin{document}
\newcommand{\figcol}[1]{{\color{blue}{#1}}}
\newcommand{\etal}{\textit{et al.}}
\renewcommand{\Re}{\text{Re}}
\newcommand{\EVROC}{$\text{e-VROC}^{\text{TM}}$}
\newcommand{\epsdot}{\dot{\epsilon}}
\newcommand{\lh}{\tilde{l}_{h}}
\newcommand{\lv}{\tilde{l}_{v}}

\title{\Huge{Optimised hyperbolic microchannels for the mechanical characterisation of bio-particles}}

\author{Yanan Liu,\textit{$^{a, e, *}$} Konstantinos Zografos,\textit{$^{c, \ddag, *}$} Joana Fidalgo,\textit{$^{a,b}$} Charles Duch{\^e}ne\textit{$^{a}$}, Cl{\'e}ment Quintard,\textit{$^{a, f}$}\\ Thierry Darnige,\textit{$^{a}$}  Vasco Filipe,\textit{$^{d}$} Sylvain Huille,\textit{$^{d}$} Olivia du Roure,\textit{$^{a}$} M\'{o}nica S. N. Oliveira,\textit{$^{b}$} Anke Lindner\textit{$^{a}$}\\ 
\hspace{1cm}\\ 
\textit{$^{a}$~Laboratoire PMMH-ESPCI Paris, PSL Research University, 10, rue Vauquelin, Paris, France.}\\ 
\textit{$^{b}$~James Weir Fluids Laboratory, Department of Mechanical and Aerospace Engineering,}\\ 
\textit{University of Strathclyde, G1 1XJ Glasgow, United Kingdom.}\\ 
\textit{$^{c}$~School of Engineering, University of Liverpool, Brownlow Street, Liverpool L69 3GH, UK.}\\ 
\textit{$^{d}$~Sanofi Biopharmaceutics, Vitry-sur-Seine, France.}\\ 
\textit{$^{e}$~School of physics, Northwest University, Xi'an, China.}\\ 
\textit{$^{*}$~YL and KZ contributed equally to this work.}\\ 
\textit{$\ddag$~current address Univ. Grenoble Alpes, CEA, Leti, DTBS, F-38000 Grenoble, France}}
\maketitle

\begin{abstract}
The transport of bio-particles in viscous flows exhibits a rich variety of dynamical behaviour, such as morphological transitions, complex orientation dynamics or deformations. Characterising such complex behaviour under well controlled flows is key to understanding the microscopic mechanical properties of biological particles as well as the rheological properties of their suspensions. While generating regions of simple shear flow in microfluidic devices is relatively straightforward, generating straining flows in which the strain rate is maintained constant for a sufficiently long time to observe the objects' morphologic evolution is far from trivial. In this work, we propose an innovative approach based on optimised design of microfluidic converging-diverging channels coupled with a microscope-based tracking method to characterise the dynamic behaviour of individual bio-particles under homogeneous straining flow. The tracking algorithm, combining a motorised stage and microscopy imaging system controlled by external signals, allows us to follow individual bio-particles transported over long-distances with high-quality images. We demonstrate experimentally the ability of the numerically optimised microchannels to provide linear velocity streamwise gradients along the centreline of the device, allowing for extended consecutive regions of homogeneous elongation and compression. We selected three test cases (DNA, actin filaments and protein aggregates) to highlight the ability of our approach for investigating the dynamics of objects with a wide range of sizes, characteristics and behaviours of relevance in the biological world.
\end{abstract}


\section{Introduction}
\label{sec_Introduction}

\noindent
The ability to characterise bio-molecule and bio-particle behaviour under transport is critical to understand biological processes. Here, we focus on the characterisation of the dynamical behaviour (e.g. evolution of morphology or conformation) of individual bio-particles under straining flow generated as a result of velocity gradients along the streamwise direction. Elongation and compression flows (referring here to flows in which fluid elements and objects are subjected to positive and negative strain rates, $ \dot{\varepsilon}$, respectively) are ubiquitous in physiological flows, as well as in bio-medical applications \cite{haward2016microfluidic}. For example, straining flows are generated as blood flows through a stenosis, or when streams of samples or reagents come together in lab-on-chip devices for testing and diagnosis, or as a fluid transitions from a syringe to a needle. Understanding the behavior of bioparticles in such flows is thus highly relevant from a fundamental point of view but also to design efficient technological applications.

\par
Microfluidic devices have been widely employed to study the dynamics of microscopic objects under flow \cite{Mai2012,Lu2017,Schroeder2018} as they present a range of advantages, including the precise control of the flow conditions, the ability to achieve high strain rates under low inertia, as well as the consumption of a small amount of sample \cite{MSNOliveira2011}. Arguably, the two most popular microfluidic designs for generating extensional flows are the cross-slot \cite{haward2016microfluidic,galindo2013microdevices} and the contraction-expansion \cite{McKinley2009,galindo2013microdevices}. 

\par
The cross-slot has two opposing inlets and two opposing outlets and requires the control of multiple streams. It has been used to characterise the elongational viscosity of polymer melts and suspensions \cite{Fischer2011,haward2016microfluidic}, to stretch cells \cite{Cha2012} and measure its viscoelastic properties \cite{Guillou2017}, and to study individual bio-polymer (e.g. DNA) dynamics \cite{McKinley2009,Velve2010} under extension. It relies on trapping molecules/particles at the stagnation point generated at the centre of the geometry. It is able to generate high strain rates, but it is well-known that in conventional devices, such as those used for investigating the behaviour of DNA \cite{hsiao2017direct} and RBC \cite{Guillou2017}, the strain rate peaks at the stagnation point but decays quickly away from it \cite{Haward2012,Haward2012b,hsiao2017direct}. These platforms are also not ideal for investigating particle behaviour under compression and maintaining molecules/particles at the stagnation point requires a delicate balance of pressure and flow rate, limiting in most cases the particle residence time \cite{kantsler2012fluctuations,de2014mechanical,hsiao2017direct}. Furthermore, the fact that the objects are trapped at the stagnation point limits the possibility to investigate the behaviour under transport and the importance of history effects. 

\par
Contraction-expansion flows involve a change in geometry and are in general simpler to operate as they require the control of a single stream. The most common designs are those with a sudden (or abrupt) contraction and/or expansion. These have been widely considered for investigating the mechanisms of fluid elasticity in artificial and natural polymer solutions both numerically  \cite{Hassager1988,Owens2002,Oliveira2003,Alves2003b,Afonso2011a} and experimentally \cite{Boger1987,Rothenstein2001,Rodd2005,Rodd2007,oliveira2007viscous}. Abrupt \cite{Gulati2008} and quasi-abrupt \cite{Gulati2015,Tomaiuolo2011,zeng2014mechanical,mancuso2017stretching} (gradual) contractions have also been employed to stretch and characterise DNA solutions \cite{Shrewsbury2001,Mai2012}, to study the deformability and velocity of healthy and artificially impaired red blood cells (RBCs) \cite{zeng2014mechanical,mancuso2017stretching,Boas2018} and to measure mechanical properties (e.g. elastic and Young's moduli) of capsules \cite{park2013transient,Chen2017,villone2019design,Goff2017}. Although abrupt contractions are easy to design, they come with some significant drawbacks. Like in the cross-slot, the straining flows they generate occur in a small region only \cite{Zografos2019}, along which the strain rate is variable, failing to produce well-controlled homogeneous extension/compression conditions. They create large jumps in flow velocity, causing difficulty in visualisation, scaling and interpretation of experimental data. Alternative geometries with tapered boundaries to create smoother transitions have also been considered. Although these empirically designed modifications reduce the problems experienced in sudden geometries, they are far from ideal in terms of achieving homogeneous extension \cite{mancuso2017stretching}.

\begin{figure}[!h]
\centering
\begin{overpic}[width=0.99\columnwidth]{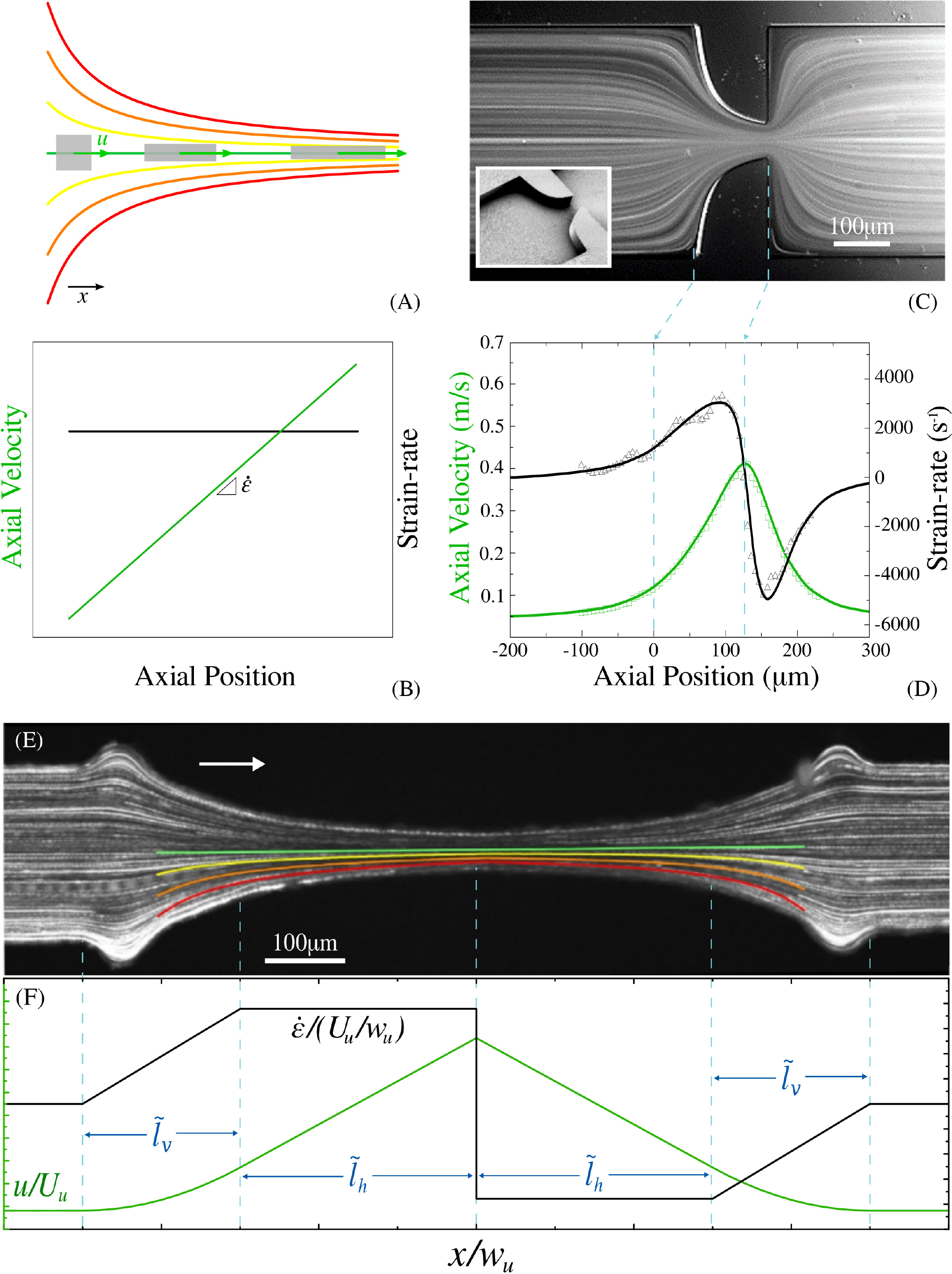}
\end{overpic}
\caption{Ideal extensional flow (A, B) in contrast with real flow behaviour in a planar hyperbolic microfluidic geometry (C, D), illustrating the need for an optimised geometry (E, F): (A) Deformation of an object in an ideal extensional flow with characteristic hyperbolic streamlines; (B) Ideal linear velocity profile (green) to generate a constant strain rate; (C) Experimentally obtained pathlines in the microfabricated channel shown in the inset (Scanning Electron Micrograph); (D) Real velocity (green) and strain rate (black) profiles obtained experimentally (open markers) and numerically (solid lines) at the centreline in the corresponding hyperbolic-shaped contraction, highlighting the deviations from the ideal profile due to entrance/exit and wall effects \cite{oliveira2007viscous}; (E) Flow patterns in an optimised geometry used in this work, showing the experimental pathlines (black and white) overlapped with a family of hyperbolae (coloured lines), highlighting the hyperbolic nature of the streamlines near the centreline; and (F) corresponding velocity (green) and strain rate profiles (black) along the centreline, where important geometrical parameters used in the optimisation are shown.}
\label{fig_1motivation}
\end{figure}

An alternative configuration exhibiting hyperbolic-shaped walls, was proposed with the intent to produce a nominal strain rate along the flow centreline \cite{Gogswell1978, James1990}. At the microscale, these geometries have been used for stretching DNA molecules \cite{Randall2006}, investigating the deformability of RBCs \cite{Lee2009,MSNO2013,Faustino2014,Faustino2019}, as rectifiers \cite{Sousa2010,Sousa2012} to study their viscoelastic behaviour of polymer solutions \cite{MSNO2007b,Wang2011} and for rheological characterisation \cite{Sousa2011,Deano2011,Ober2013,McKinley2016}. Despite their remarkably better performance relative to the abrupt and tapered designs, they still suffer from non-homogeneous entry and exit effects \cite{oliveira2007viscous,Ober2013,Zografos2019}.
 
\par
Even though mathematically a pure straining flow corresponds to a hyperbolic flow (\cref{fig_1motivation}\figcol{A}), fabrication of a hyperbolic shaped channel (e.g. \cref{fig_1motivation}\figcol{C}) is not sufficient to obtain such a flow. This is due, on the one hand, to the no slip boundary condition at the walls of microfluidic channels and, on the other hand, to the continuously changing aspect ratio of the channel along the contraction in the case of planar geometries of constant depth typical of microfluidics. This is clearly shown in \cref{fig_1motivation} where we contrast the ideal flow  (\cref{fig_1motivation}\figcol{A}, \figcol{B}) with the flow actually obtained in a hyperbolic microchannel (\cref{fig_1motivation}\figcol{C}, \figcol{D}) \cite{oliveira2007viscous}. The deviation from the ideal linearly increasing velocity profile and corresponding constant strain rate highlights the need for device optimisation.

\par
The microfluidic flow geometries used in this work were optimised using the numerical procedure discussed in \citepp{zografos2016microfluidic} to provide relatively wide regions of uniform (positive and negative) strain rates. A realisation of such an optimised microfluidic channel is shown on \cref{fig_1motivation}\figcol{E} together with experimental pathlines. A family of hyperbolae is also represented and demonstrates good agreement with the experimental pathlines close to the centreline. Indeed, a linear velocity profile and constant strain rates are obtained over an extended region (\cref{fig_1motivation}\figcol{F}). We use the optimised microfluidic devices to assess the dynamics of bio-particles in homogeneous straining flows. To this means, we developed a specific tracking algorithm to follow the bio-particles during transport to obtain good quality images throughout the channel.

\par
To demonstrate the wide domain of application of our microfluidic device, we have first chosen two relevant bio-polymers with very different properties: actin filaments and Deoxyribonucleic acid (DNA) molecules. DNA is a semi-flexible biopolymer that can be found in humans and almost all other living organisms, which carries genetic information crucial for their development, functioning, growth and reproduction. At equilibrium, thermal entropic forces favour coiled configurations of typical sizes up to a few microns, whereas the polymer contour length can be as large as several tens of microns. DNA is considered a model semi-flexible biopolymer, and its microscopic dynamics under flow have been intensely studied \cite{Schroeder2018} and related to the shear-dependent viscosity and normal stress differences, leading to rod climbing and Weissenberg effects in solutions of long polymers \cite{Bird1987,Doi1986}. Actin filaments are flexible Brownian filaments abundant in cells, where they form different kinds of dynamic structure like actin cortex, lamellipodium or stress fiber, serving different functions including cell shape and cell motility \cite{Pollard2003,Pollard2009}, cell division \cite{Garner2007}, cell signaling \cite{Dustin2000,Calderwood2000}, as well as the establishment and maintenance of cell junctions \cite{Baum2011}. At rest, in vitro, they behave as straight shape-fluctuating filaments of typical length of tens of microns. Their microscopic dynamics have recently been investigated under shear and straining flows \cite{liu2018morphological,Chakrabarti2020}. The dynamics of an individual flexible filament in viscous flows underlies a wealth of biophysical processes from flagellar propulsion \cite{Lauga2009} to intracellular streaming \cite{Shelley2016} and are the key to deciphering the rheological behaviour of many complex fluids and soft materials \cite{Roure2019}.  A third example studied here is a protein aggregate, which can be formed in solutions of monoclonal antibodies \cite{Engelsman2011}. These aggregates can cause immunogenicity \cite{Ahmadi2015} and represent a threat to the reliability of injection devices \cite{Rathore2012}. Understanding aggregate transport dynamics is thus crucial for numerous biomedical applications and in particular for the prevention of clogging of syringes during injection processes \cite{Duchene2020}. The size and properties of the aggregates studied here are distinctly different from the two bio-polymers, with sizes of around hundred microns and rather compact shapes. Discussing these three selected test objects we highlight the ability of the design and techniques proposed for investigating the dynamics under extension and compression in very well defined flow conditions of objects with a wide range of sizes, characteristics and behaviour of relevance in the biological world.
 
\par
The remainder of the paper is organised as follows. In section \nameref{sec_channeldesign}, the numerical techniques used to optimise the shape of the device are summarised, the channel design is discussed and its performance is compared against commonly employed configurations. In section \nameref{sec_implementation} we detail the experimental set-up, fabrication protocols and techniques used for flow and particle characterisation. Section \nameref{ResDisc} describes the observations made on the microscopic dynamics of the three bio-particles under extension and compression and the main conclusions are summarise in section \nameref{Conc}.
%
%
\section{Geometry design}
\label{sec_channeldesign}
\noindent
In this section, we discuss the procedure followed for designing the optimised converging-diverging geometries used in this study. We employ optimised 3D microfluidic contraction-expansion devices of constant depth, which are designed to exhibit a region of homogeneous strain rate along the flow centreline \cite{zografos2016microfluidic}. These optimised shapes have been shown to have enhanced performance relative to the hyperbolic geometries \cite{zografos2016microfluidic}, in particular for some aspect ratios and short contraction lengths when the hyperbolic acts almost as an abrupt contraction \citep{Zografos2019}. Here, we compare its performance to that of more conventional designs, such as the sudden expansion and tapered configurations.

\par
The numerical procedure is based on an iterative optimisation strategy combining an automatic mesh generation routine, a computational fluid dynamics solver and an optimiser. The CFD simulations performed at every single optimisation step, consider a laminar, incompressible and isothermal, Newtonian fluid flow. The continuity and the Navier-Stokes equations are discretised in a finite volume framework and are solved numerically:

\begin{equation}
\nabla\cdot\pmb{\text{u}} = 0, 
\label{eq_cont}
\end{equation}
\begin{equation}
\rho \left(\frac{\partial{\pmb{\text{u}}}}{\partial{t}} + \pmb{\text{u}}\cdot\nabla\pmb{\text{u}}\right) = -\nabla p + \nabla \cdot \pmb{\tau},
\label{eq_mom}
\end{equation}
where $\pmb{\text{u}}$ is the velocity vector, $\rho$ is the fluid density, $p$ is the pressure, $t$ is time and $\pmb{\tau}$ corresponds to the stress tensor. The discretised set of partial differential equations, \cref{eq_cont,eq_mom}, are solved using an in-house implicit finite volume CFD solver, developed for collocated meshes \cite{Oliveira1998,alves2003convergent}. The pressure and velocity fields are coupled using the \mbox{SIMPLEC} algorithm for collocated meshes by employing the Rhie and Chow interpolation technique \cite{rhie1983numerical}.
In the simulations we apply creeping flow conditions ($\Re \rightarrow0$, which is a good approximation for the microfluidic flow conditions considered here) by setting the convective terms in the momentum equation equal zero ($\pmb{\text{u}}\cdot\nabla\pmb{\text{u}}=0$). 
\par
The optimisation procedure relies on the use of Non-Uniform Rational B-Splines (NURBS) for applying free-form deformations \cite{Lasmunsin1994}. The outline of the geometry is shaped in such a way that the velocity profile at the flow centreline, $u$, approximates the desired target velocity profile (see \cref{fig_1motivation}\figcol{F}):
\begin{equation}
   \tilde{u} = \left\{
	\begin{array}{l l}
	  \tilde{u}_{u}                  
	  &\quad \text{if} \quad \tilde{x} \le -\lh-\lv  \\
	  f_{2} \left[ \tilde{x} + \lh+\lv \right]^{2}+\tilde{u}_{u}   
	  &\quad \text{if} \quad -\lh-\lv\le \tilde{x} \le -\lh  \\
	  \hspace{8pt} f_{1} \tilde{x}  + \tilde{u}_{c}  
	  &\quad \text{if} \quad -\lh \le \tilde{x} \le 0 \\
	  -f_{1} \tilde{x} + \tilde{u}_{c}  
	  &\quad \text{if} \quad 0 \le \tilde{x} \le \lh  \\
	  f_{2}\left[ \tilde{x} -\lh-\lv\right]^{2} + \tilde{u}_{u}   
	  &\quad \text{if} \quad \lh \le \tilde{x} \le \lh + \lv  \\
	  \tilde{u}_{u}
	  &\quad \text{if} \quad \tilde{x}\ge \lh + \lv \\
        \end{array}\right.
        \label{eq_profile_U}
\end{equation}
All symbols with ``tilde'' are used to represent normalised quantities, with all lengths normalised using the upstream width $w_{u}$ as reference and velocities with the average velocity in the upstream straight channel $U_{u}$ (e.g. $\tilde{x}=x/w_{u}$, $\tilde{u}=u/U_{u}$), to increase their readability. The subscripts 'c' and 'u' refer to the throat of the contraction and the upstream region, respectively. The resulting normalised strain rate, $\dot{\varepsilon}=\partial u / \partial x$ is shown in \cref{fig_1motivation}\figcol{F} and is given by:
\begin{equation}
   \dot{\varepsilon}/(U_{u}/w_{u}) = \left\{
	\begin{array}{l l}
	  0                  &\quad \text{if} 
	  \quad \tilde{x}\le -\lh-\lv  \\
	  2f_{2}[ \tilde{x} + \lh+\lv]   
	  &\quad \text{if} \quad -\lh-\lv\le \tilde{x} \le -\lh  \\
	  f_{1}  
	   &\quad \text{if} \quad -\lh \le \tilde{x} \le 0 \\
	  -f_{1}  
	  &\quad \text{if} \quad 0 \le \tilde{x} \le \lh  \\      
	  2f_{2}[ \tilde{x} -\lh-\lv]   
	  &\quad \text{if} \quad \lh \le \tilde{x} \le \lh + \lv \\
	  0                  
	  &\quad \text{if} \quad \tilde{x}\ge \lh + \lv \\
        \end{array}\right.
        \label{eq_profile_SR}
\end{equation}
\par 
The optimised part of the geometry is, thus, composed of an elongation part (positive strain rate) followed by a compression part (negative strain rate). Each of these parts combines a region of homogeneous strain rate (of length $\lh$) in which the velocity is varying linearly along the flow centreline and a transition region (of length $\lv$) as shown in \cref{fig_1motivation}\figcol{F}. In this transition region the strain rate varies linearly to guarantee a smooth transition from zero in the straight part of the domain to the desired value in the region of interest as shown in the profiles of \cref{fig_1motivation}\figcol{F}. In both \cref{eq_profile_U,eq_profile_SR}, the dimensionless parameters $f_{1}$ and $f_{2}$ are defined as \mbox{$f_{1}=(\tilde{u}_{c}-\tilde{u}_{u})/(\lh+\lv/2)$} and \mbox{$f_{2}=(\tilde{u}_{c}-\tilde{u}_{u})/[2\lv(\lh+\lv/2)]$}, respectively.
\par
Essentially, the optimisation procedure requires as input the definition of the contraction/expansion ratio ($CR=w_{u}/w_{c}$), the height of the channel (or the corresponding aspect ratio $AR=w_{u}/H$) and the normalised lengths $l_{h}/w_{u}$ and $l_{v}/w_{u}$, to be able to generate the shape outline of the microchannel. One important point to note about 3D planar contraction-expansions is that the local channel aspect ratio varies along the region of interest, since $H$ is constant while the width is varying. 
As a result, for geometries with high $CR$ and moderate $AR$ in particular (i.e. away from the 2D limit of $AR\rightarrow0$) the ratio of the velocity at the flow centreline to the corresponding average velocity varies along the straining region (i.e. ${u}_{u}/U_{u}$ is different from ${u}_{c}/U_{c}$), even if assuming instantaneous fully developed flow \cite{white2006viscous}. The Hencky strain estimated as $\varepsilon_{H}=\int \epsdot dt = \text{ln}(\frac{u_{c}/U_{c}}{u_{u}/U_{u}} CR)$ is thus fixed for each geometry, while the strain rate can be varied by changing the flow rate through the device. Note that when $AR << 1$, the 2D limit is a good approximation and the Hencky strain reduces to $\varepsilon_{H}\simeq \text{ln}(CR)$.
\par
Here, we present results for an optimised geometry with nominal values $CR=8$, $AR=8$, $\tilde{l}_{h}=1.5$ and $\tilde{l}_{v}=1.0$ that will be used throughout the paper. At some occasions a second optimised geometry with $CR=8$, $AR=2$, $\tilde{l}_{h}=1.5$ and $\tilde{l}_{v}=1.0$ will be used. This will be explicitly mentioned. 
\par
\begin{figure}[t]
\centering
\begin{overpic}[width=0.97\columnwidth]{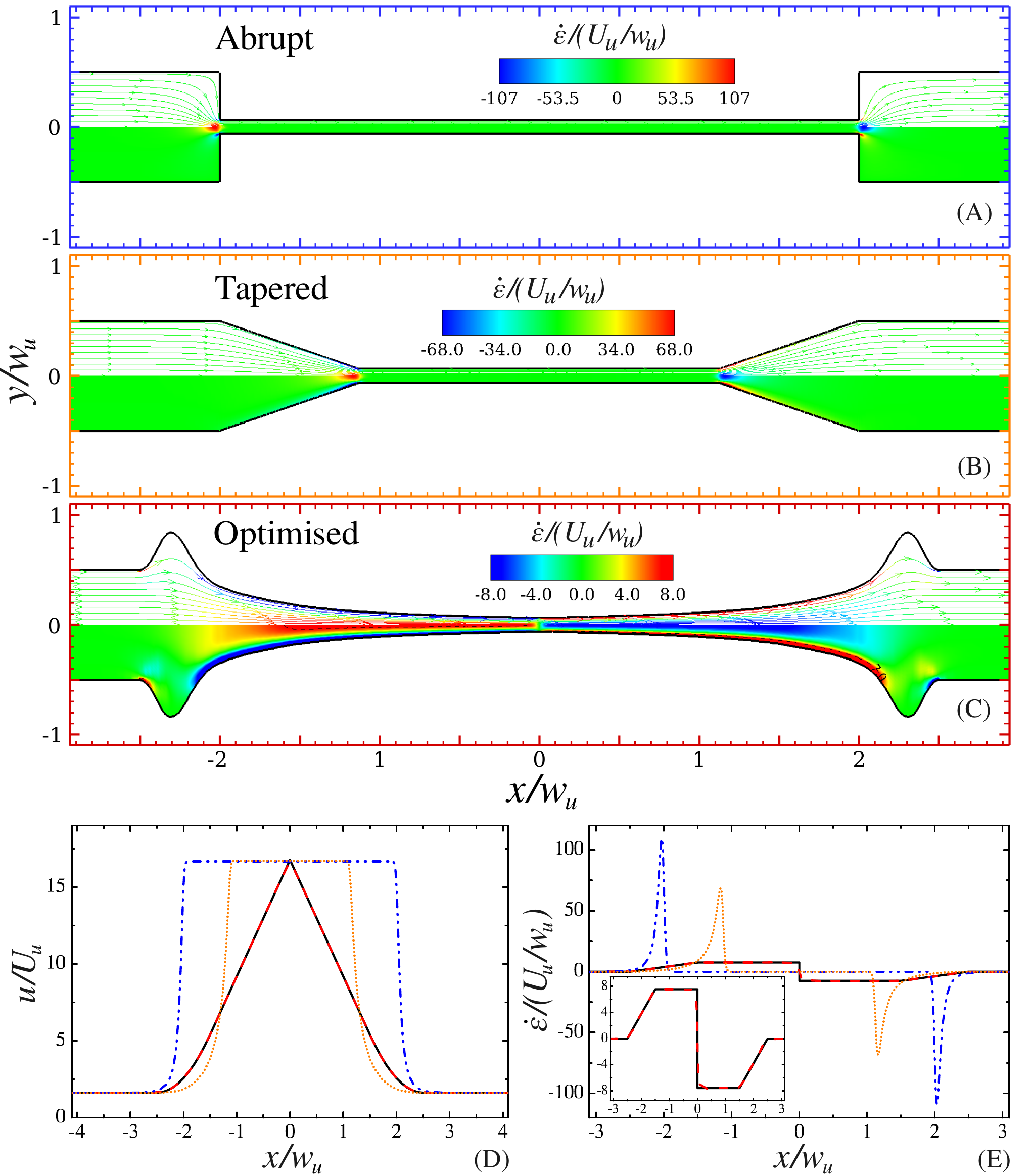}
\end{overpic}
\caption{Comparison of the flow field in (A) Abrupt, (B) $45^{\text{o}}$ Tapered, and (C) Optimised contraction-expansion geometries: (A-C) shows the normalised strain rate contour-plots (bottom half) and coloured streamlines (top half); (D-E) compares their performance in terms of (D) velocity and (E) applied strain rate profiles along the flow centreline. The lines in (D) and (E) correspond to the desired target profile (\textit{solid-line}) and the profiles generated by the abrupt (\textit{dashed/dotted-line}), the $45^{\text{o}}$ tapered (\textit{dotted-line}) and the optimised (\textit{dashed-line}) designs. The zoom view in the inset (E) demonstrates the good performance of the optimised geometry relative to the target.}
\label{fig_FIG3}
\end{figure}

To highlight the performance of the numerically optimised device relative to commonly used conventional contraction-expansion shapes (e.g. the abrupt contraction-expansion and the $45^{o}$ tapered contraction-expansion), in \cref{fig_FIG3} we compare the numerically calculated flow field in these three geometries (all with the same nominal values of $CR=8$ and $AR=8$).
The top half of \cref{fig_FIG3}\figcol{A}-\figcol{C} display the streamlines coloured by normalised strain rate, while the bottom half displays the contour plot of the normalised strain rate. The comparison of the velocity and strain rate profiles along the centreline are shown in \cref{fig_FIG3}\figcol{D} and \cref{fig_FIG3}\figcol{E}, respectively. It is clear that for the conventional designs there is only a small region of non-zero strain rate in which additionally the strain rate is not constant. In fact, the strain rate quickly peaks to values substantially higher than in the target and in the optimised geometry profiles (by approximately an order of magnitude) and then decays rapidly. Such very high strain rates, even if remaining very localised, can potentially damage fragile bio-particles or induce an ill defined deformation history on the particle that can influence the behaviour in the contraction flow. In the optimised geometry on the other hand there is a large plateau region ($1.5w_{u}$ or $12w_{c}$) of constant strain rate along the centreline.

\section{Microfluidic implementation}
\label{sec_implementation}

In this section we describe the experimental implementation of a microfluidic setup using channels with optimised dimensions as well as the tracking method that we have developed to follow particles transported along the channel permitting observation of their shape and orientation with excellent resolution. 

\subsection{Experimental set-up}
\begin{figure}[!t]
	\begin{center}
		\includegraphics[width=0.95\columnwidth]{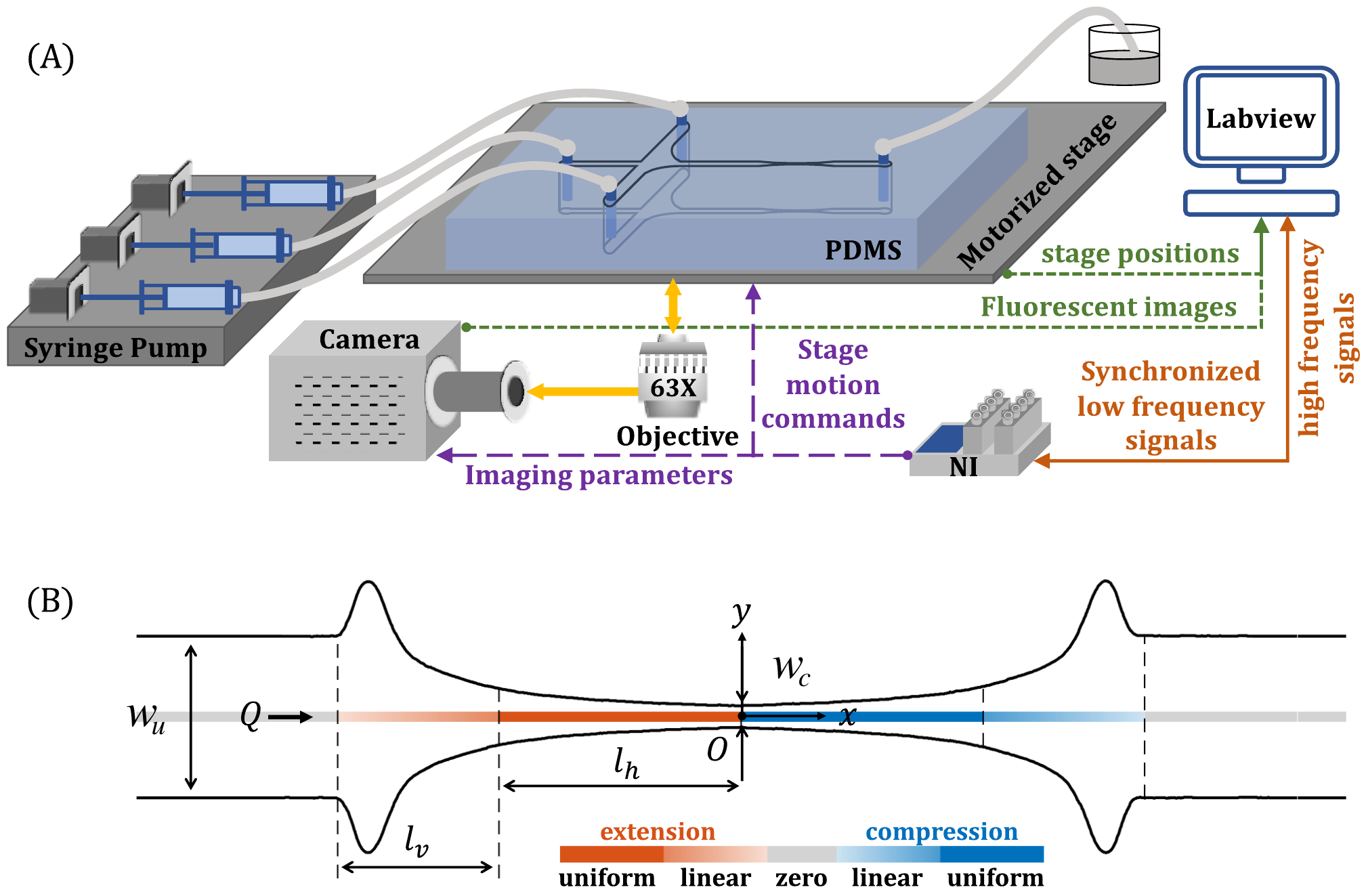}
	\end{center}
	\caption{(A): A detailed sketch of the experimental setup, showing microfluidic channel, flow control and tracking system; (B): Shape of the microfluidic channel with key geometrical parameters. The different colours represent regions with different strain rate $\dot{\epsilon}$ behaviour, see also \cref{fig_1motivation}\figcol{F}, with grey corresponding to $\dot{\epsilon}=0$, and red (blue) corresponding to positive (negative) $\dot{\epsilon}$. Colour gradient indicates linear variation of  $\dot{\epsilon}$ with distance while dark colours show region of constant  $\dot{\epsilon}$. An experimental realisation of such a channel can be seen on \cref{fig_1motivation}\figcol{E}.
}
	\label{fig_sketch}
\end{figure}

The set-up used for this purpose is shown schematically in ~\cref{fig_sketch}\figcol{A}, and includes the optimised microfluidic channel and the tracking system. A syringe pump (neMESYS 290N) for flow rate control, a fluorescence microscope (Zeiss Observer A1) equipped with a 63x objective (C-Apochromat 63x/1.20 W Corr M27) and a motorised stage (ASI PZ-2500).  Images are captured using a CMOS camera (HAMAMATSU ORCA flash 4.0LT, 16 bits) with a frequency of 10-125Hz. The total flow rates $Q$ employed vary in the range of  $3.1-5.5\,$nL/s, providing average velocities around $100\upmu$m$/$s and maximum velocities around 1~mm/s, corresponding to a range of strain rate of 0.3 $\sim$ 5$s^{-1}$ and maximal Reynolds numbers of $10^{-2}$. Inertia can thus be neglected in all our experiments.

\subsection{Microfluidic flow channel }

Microfluidic channels have been fabricated using standard soft lithography techniques following the design of the optimised geometry for $AR=8$, as described in \cref{sec_channeldesign}. The height of the channel $H$ is fixed to be $100\pm1\upmu$m, with a minimum width in the contraction $w_c=100\pm 5 \upmu$m and an upstream width $w_u=800\pm 10\upmu$m. The optimised straining part of the device along which the widths of the channel is changing is composed of the transition region with linearly increasing strain-rate of length $l_v=w_u$ and the region of homogeneous strain-rate of length $l_h=1.5 w_u$ (see \cref{fig_sketch}\figcol{B}). The total length of the optimised region, including the extension (highlighted in red) and the compression (highlighted in blue) regions is thus $2(l_v+l_h)$. A realisation of this microchannel can be seen on \cref{fig_1motivation}\figcol{E}.

Since strain rates have been optimised only for the centreline of the channel, particles to be analysed should be transported along this line. For this purpose, we used a flow focusing with three identical branches well upstream of the hyperbolic channel. When the flow rates of the outer branches are 10 times larger than that of the middle branch, the particles transported in the latter are confined in the region close to the centreline\cite{oliveira2012divergent}.

\subsection{Real flow profile and strain rates}
\label{subsec:ptv}
\par
\begin{figure}[t!]
	\begin{center}
		\includegraphics[width=0.99\columnwidth]{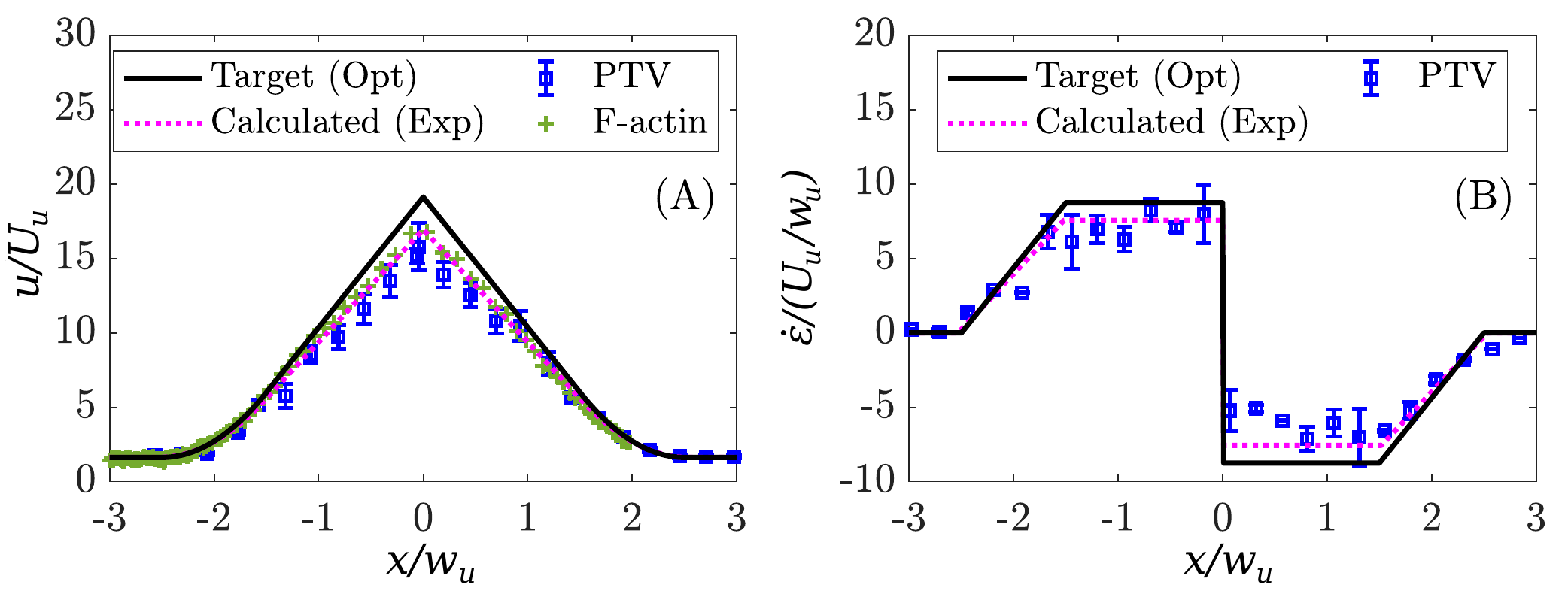}
	\end{center}
	\caption{Normalised velocity profiles (A), and corresponding strain rate profiles (B) along the central streamline in the optimised channel. The black solid lines represent the target profiles, which overlap with the numerically calculated profiles along the centreline of the ideal optimised geometry. All other sets refer to data obtained in the real fabricated channel, which deviates slightly form the optimised shape due to limitations inherent to the fabrication procedure. The pink dotted lines represent the calculated profiles, the blue squares represent the profiles obtained using PTV at fixed positions in the channel, and the green crosses represent the velocity experienced by a bio-particle, in this case an actin filament, transported along the centreline tracked using the proposed tracking method.	}
	\label{fig_PIV}
\end{figure}

Due to the limited resolution of the microfluidic fabrication techniques used, the channel dimensions deviate from the ideal optimised dimensions within the error bars given in the previous Section. To account for this difference we have extracted the full shape of the channel using image treatment procedures and have calculated the resulting flow profile. \Cref{fig_PIV} shows a comparison between profiles calculated for the optimised shape and for the real experimental shape (black and magenta lines, respectively). The real channel dimensions lead to a slightly smaller value of $\dot{\epsilon}$ for identical average flow velocities compared to the optimised channel, however the main features of the optimisation procedure hold and we still obtain a large region with constant and well controlled extension and compression rates. 
\par
In order to examine experimentally the velocity profile in the hyperbolic channel, we used particle tracking velocimetry (PTV). The results are shown as  blue points in figure~\cref{fig_PIV} for a flow rate of $Q=5.5$nl/s.  Particle tracking velocimetry is carried out with a dilute suspension of fluorescent particles, with a diameter of $1 \upmu$m at a series of fixed positions. Images are captured with a frequency of 50-125Hz.  Note that both geometries with and without flow focusing have been used and identical results have been obtained. The good agreement between the calculated velocity profile and the particle tracking velocimetry shows the good control of a stable flow field with our set-up and the flow focusing device.

\subsection{Particle tracking}
\label{subsec_implementation}

The simplest way to observe particles in straining flows consists in taking pictures at fixed locations of the channel with a fast camera and  a short exposure time~\cite{mancuso2017stretching,strelnikova2017direct}. A sufficiently short exposure time is necessary to prevent image blur of the particle as it is transported during the exposure. Bio-particles with nanoscale cross-sections, like actin filaments or DNA molecules are not visible in bright field microscopy and usually require fluorescent imaging with rather long exposure times of up to fifty milliseconds to be observed. Furthermore, high magnification objectives are necessary to resolve their shape but  limit the observation field. Hence, for good image quality it is necessary to follow particles while transported in the flow, for example by the use of a motorised stage \cite{Fidalgo2017}.

Here, we opt to observe particles transported along the centre streamline in the mid-height of optimised channels. It is very demanding to track particles transported in strongly accelerating and decelerating velocity fields as those in converging-diverging channels. In our particular example, the velocity varies from $100\upmu$m/s in the straight parts of the channel up to 1~mm/s at the throat of the constriction. Note that particle tracking requires two conditions to be fulfilled. First, the stage needs to follow as closely as possible the particle transported in the optimised channel to keep it in the field of view. Second, the velocity of the stage movement and the  velocity of the particle need to be as identical as possible at every instant to avoid image blur during the long exposure time. 

To fulfil the two conditions, we developed a tracking system using a Labview program and a motorised stage, synchronising stage movement with image capture as depicted in ~\cref{fig_sketch}\figcol{A}. The red line shows that high frequency signals ($10^6$ Hz) are translated into two synchronised low frequency signals, 10Hz and 20Hz respectively, by a PC equipped with a Labview program. Through the NI I/O device, the two synchronised signals are sent to the camera and the stage (purple lines), triggering read out of the stage position and image capture (green lines) and transmitting the stage velocity.

\begin{figure}[t!]
	\centering
	\includegraphics[width=0.99\columnwidth]{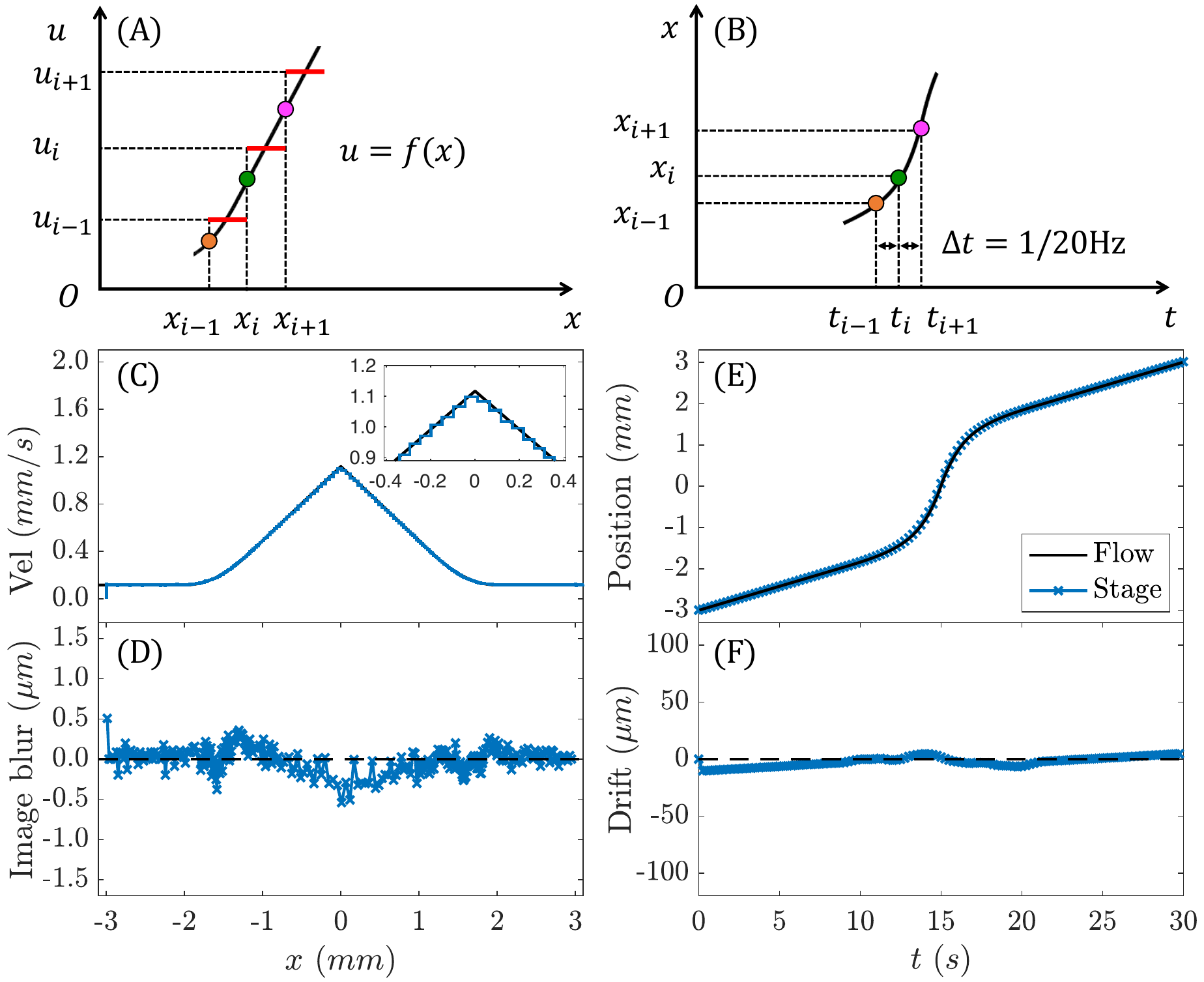}
	\caption{Tracking method and accuracy. (A) and (B) schemes illustrate how velocity commands to be sent are calculated. Discrete stage velocity $u_i$, red horizontal line, is calculated as $(x_{i+1}-x_i)/ \Delta t$, according to returned $x_{i-1}$ and known relation between position and time. Followed with the comparison between flow profile (black solid lines) and actual stage movement (blue markers). (C): velocity as a function of position; (D) displacement as a function of time; (E): Image blur due to velocity difference during exposure time as a function of position; (F) particle drift due to the difference between displacements as a function of time.}
	\label{Fig:Blur}
\end{figure}

The algorithm controlling the stage velocity uses the actual stage position to send velocity commands at fixed intervals $\Delta t=1/20s$ during which the stage velocity is constant. However, during such an interval the flow velocity is not constant and the stage velocity thus needs to be carefully chosen to reproduce identical displacement of stage and particle transport during each step. To evaluate the command to be sent, we use the velocity profile calculated considering the real dimensions of the channel (\cref{fig_PIV}, black line) that relates the position in the channel ($x$) to the velocity of the flow ($u$) (see \cref{Fig:Blur}\figcol{A}). At each time step $t_i$, the known stage position $x_{i-1}$ is used to infer the position $x_i$ and $x_{i+1}$ of the stage at $t_i$ and $t_{i+1}$, respectively (see \cref{Fig:Blur}\figcol{B}), from which the command $u_i$ is determined using $u_i =(x_{i+1}-x_i)/\Delta t$. Fine tuning of the parameters has been performed through a Mathematica code to further improve the performance of the algorithm, by taking into account additional delays in the communication between hardware and software and the discrete nature of the stage velocities. This leads to a very good agreement both for the velocities and the positions (see~\cref{Fig:Blur}\figcol{C, D}). The position difference is less than 20 $\mu m$ smaller than the size of the field of view of around $200\mu m$, ensuring that the particle will stay in the observation field  (see \cref{Fig:Blur}\figcol{F}). The difference in velocity between the stage and the particle is also small enough to limit the blur of the object during the 50 ms exposure time to only a fraction of microns (see \cref{Fig:Blur}\figcol{E}), i.e. much smaller than the object length. This powerful method enables us to accurately follow the objects during all the phases of the flow velocity variation (described in \cref{fig_sketch}\figcol{B} with images of very good quality giving access to the precise shape (orientation and deformation) of the object while submitted to known viscous stresses. Using our experimental set-up the maximum velocities in the throat for which our tracking algorithm fulfills the above mentioned criteria is 1~mm/s.

We first tested our tracking algorithm on individual actin filaments, and determined their local velocity. The obtained result is shown on \cref{fig_PIV} as green crosses and is in good agreement with the PTV data (blue crosses) obtained at given positions in the channel with an immobile stage confirms the high reliability of our tracking system. These results  show in addition that the presence of the bio-particle does not disturb the flow profile in a significant way. 

\subsection{Bio-particle fabrication protocols}

In this section we describe the procedure to prepare the three different bio-particles used in this work. 

\textit{$a.$ Fluorescent actin filaments.}
Concentrated G-actin, which is obtained from rabbit muscle cells and purified according to the protocol described in~\cite{spudich1971regulation}, 
is placed into F-Buffer (10$\,$mM Tris-Hcl pH$=$7.8, 0.2$\,$mM ATP, 0.2$\,$mM CaCl$_2$, 1$\,$mM DTT, 1$\,$mM MgCl$_2$, 100$\,$mM KCl, 0.2$\,$mM EGTA 
and 0.145$\,$mM DABCO) at a concentration of 1$\,\mu$M. At the same time, Alexa488, Fluorescent Phalloidin in the same molarity as G-actin is added to 
stabilise and label actin filaments. After 1 hour of spontaneous polymerisation in the dark at room temperature, concentrated F-actin solution is diluted $20\sim50$ times 
for the following experiments. 45.5\%(w/v) sucrose is added to match the refractive index of the PDMS channel ($n=1.41$) in order to get better image contrast. The viscosity of the suspensions with 45.5\%(w/v) sucrose is $5.6\,$mPa$\cdot$s at $24$ $^\circ$C, measured by an rheometer Anton Paar MCR 501.

\textit{$b.$ Fluorescent DNA molecules.}
T4 GT7 DNA (169 kbp, NIPPON GENE) is used and dyed with one YOYO-1 per four DNA base pairs followed a previous protocol~\cite{fidalgo2019microfluidics}. The 0.04ppm DNA solutions are prepared by mixing 30 $\mu$L concentrated primary T4 DNA solution (1.4 $mg/$L, 10mM pH=10 Tris-HCl Buffer) with 3.25 $\mu$L 5 $\mu$M YOYO-1 solution, and 50 $\mu$L beta-mercaptoethanol 4\% (v/v), as well as 15 $\mu$L Oxygen scavenger solution (0.2mg/mL $\alpha$-D-glucose, 0.47 mg/mL glucose oxidase, 0.16 mg/mL catalase) to reduce photobleaching. This final solution is gently rotated at 10 rpm in a dark environment for around $4\sim6$ hours to insure sufficient contact and combination between YOYO-1 and DNA molecules. 45.5\%(w/v) sucrose is also added to match reflection index thus resulting in the solvent viscosity to be $5.6\,$mPa$\cdot$s.

\textit{$c.$ Protein aggregates.}
An IgG4 monoclonal antibody (mAb) solution was provided by Sanofi (Vitry-sur-Seine, France) and formulated at $150$mg/mL in histidine buffer at pH $6.1$ and additional stabilising excipients. Aggregates were then created in the laboratory by applying heat stress consisting of incubating the mAb at $60^\circ$C for $35$min in a heating block Thermostat C (Eppendorf, Hamburg, Germany).  A large variety of different aggregate sizes, shapes and density can be produced in this way~\cite{Duchene2020}. Here, we focus on aggregates of sizes $50-100 \mu $m. The stressed samples were kept at $4^\circ$C until further use. Dilution with unstressed antibody solution was performed until a final solution viscosity of 13mPas was reached. More details on the procedure can be found in the previous work~\cite{Duchene2020}.



\section{Results}
\label{ResDisc}
\noindent
In this section, we demonstrate the performance of optimised channels for the analysis of bio-particles morphologies transported in extension and compression flows by discussing several examples. Different biopolymers as well as protein aggregates have been observed during their transport in our channels. These studies demonstrate how the combination of perfectly controlled strain rates with very precise observation capabilities allows for unprecedented qualitative as well as quantitative analysis of bio-particles dynamics under transport. 

\subsection{Extension and compression of biopolymers}

\begin{figure}[!b]
	\begin{center}
		\includegraphics[width=0.95\columnwidth]{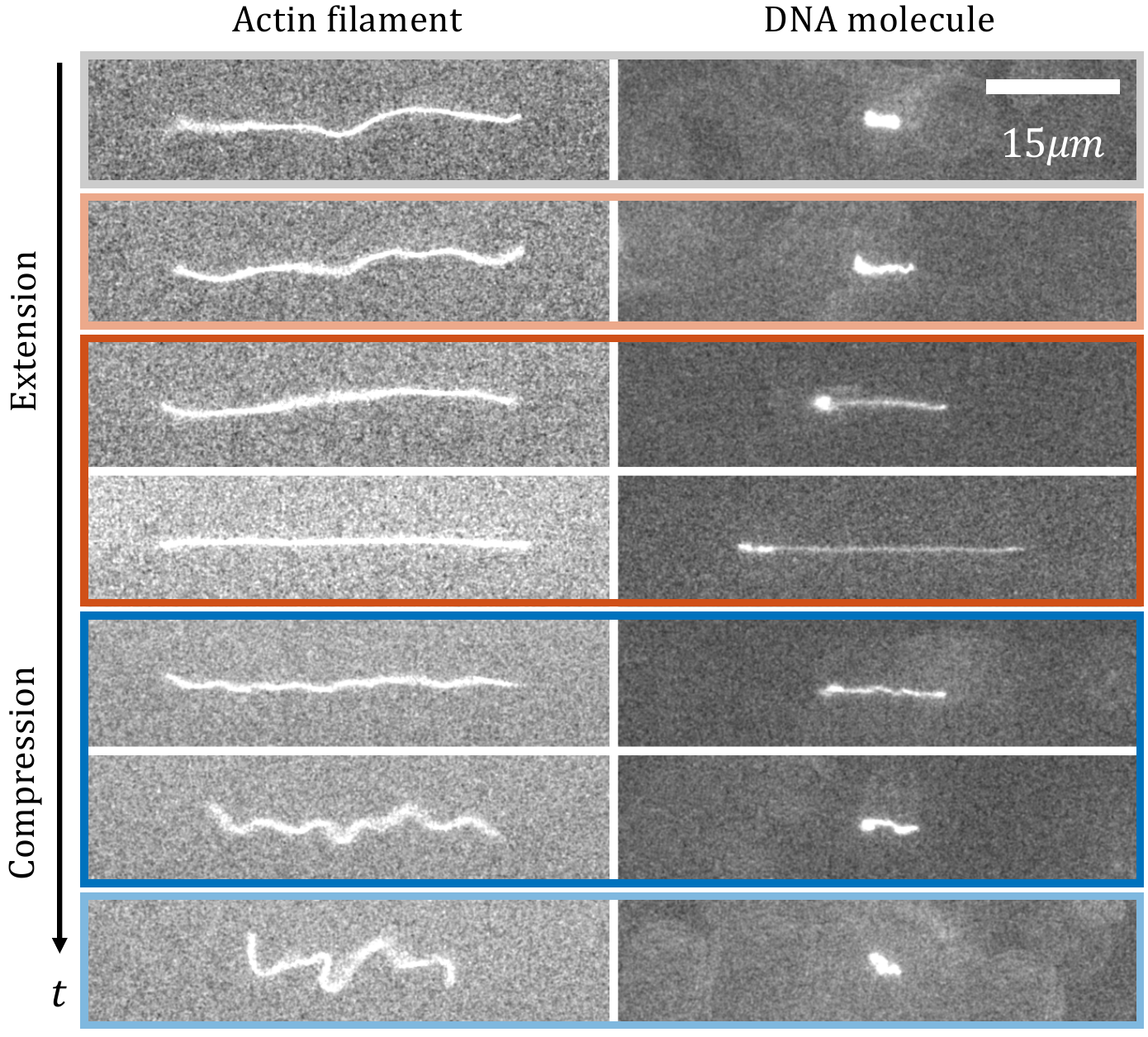}
	\end{center}
	\caption{Series of fluorescent images of two types of semi-flexible filaments in a straining flow: actin filament ($L_c=42\mu m$) and T4 GT7 DNA molecule ($L_c=72\mu m$). The colour of each frame corresponds to the region in the channel following the colour code of  figure~\cref{fig_sketch}\figcol{B}.}
	\label{fig_DNA_strain}
\end{figure}

The optimised converging and diverging channels provide a succession of extension and compression flow and thus constitute a  versatile tool for the study of deformations of different types of polymers induced by viscous forces. Here, we investigate two biopolymers that differ by their ratio of persistence length to contour length. Stabilised actin filaments have a persistence length of $\ell_p=17\pm1\mu m$ and contour length varying between $L_c=5-50\mu$m~\cite{liu2018morphological}. In a quiescent fluid, they experience bending fluctuations induced by thermal noise, but in average have a straight conformation. On the other hand, YOYO-1 dyed T4 DNA molecules have slightly larger contour length, $L_{c}\approx72\mu$m, but a much smaller persistence length of $\ell_p=57$nm favouring, due to thermal entropic forces, a coiled conformation at rest with a radius of gyration $R_g=2.7 \mu m$ \cite{Schroeder2018} \footnotemark.
Note that the presence of YOYO-1, which is DNA intercalant, increases the contour length of the DNA as compared to the same undyed molecule\citep{perkins1995stretching,thamdrup2008stretching}, whereas AFM measurements show no appreciable effects of YOYO-1 on the persistence length $\ell_p$\citep{kundukad2014effect}.
Moreover, in a dilute solution with a solvent viscosity of 5.6mPa$\cdot$s, as in our case, the relaxation time of T4 DNA is estimated to be $\lambda\approx 4.3s$~\cite{liu2009concentration}.

\footnotetext{The radius of gyration of T4 DNA in bulk reads $R_g=(\ell_pw_eL_c^3)^{1/5}$. In this expression, $w_e$ is the effective width of DNA which is highly dependent on the monovalent salt content and in our case ($10$mM NaCl solution), $w_e=8$nm ~\citep{vologodskii1995modeling}. As well using $\ell_p=57 $nm and $L_c=72 \mu $m, we are able to calculate the radius of gyration of the dyed DNA to be $2.7\mu m$~\citep{thamdrup2008stretching,schaefer1980dynamics}.}

Completely different dynamics are observed for these two types of polymers in the converging and diverging channels as shown in figure~\cref{fig_DNA_strain} where left/right columns correspond to actin filaments/DNA molecules. In the straight part of the channel (labeled in grey, top row), where the strain rate is negligible, the difference between the conformations at rest is clearly visible: the actin filament appears as a straight shape fluctuating filament, whereas the DNA molecule adopts a more compact coiled-conformation. 
During transport in the channel, the dynamic evolution of its morphology results from the interplay between viscous drag forces applied by the flow and the thermal entropic forces favouring coiled conformations. In the extension part, labelled in red, the viscous forces pull on the polymers. In the case of the actin filament this leads to the suppression of the transverse fluctuations, whereas the DNA molecule undergoes a radical shape change and transits from a coiled to a stretched conformation, known as the coil-stretch transition~\cite{de1974coil}. Both polymers appear nearly straight when reaching the throat of the constriction (fourth row of \cref{fig_DNA_strain}). In the compressive part (labeled in blue,  bottom half of \cref{fig_DNA_strain}), the DNA molecule is pushed back to its coiled equilibrium conformation, whereas the long actin filament undergoes a transition to a three-dimensional compact shape, corresponding to a buckling instability as we recently showed~\cite{chakrabarti2020flexible}.

\begin{figure}[!b]
	\begin{center}
	\includegraphics[width=0.9\columnwidth]{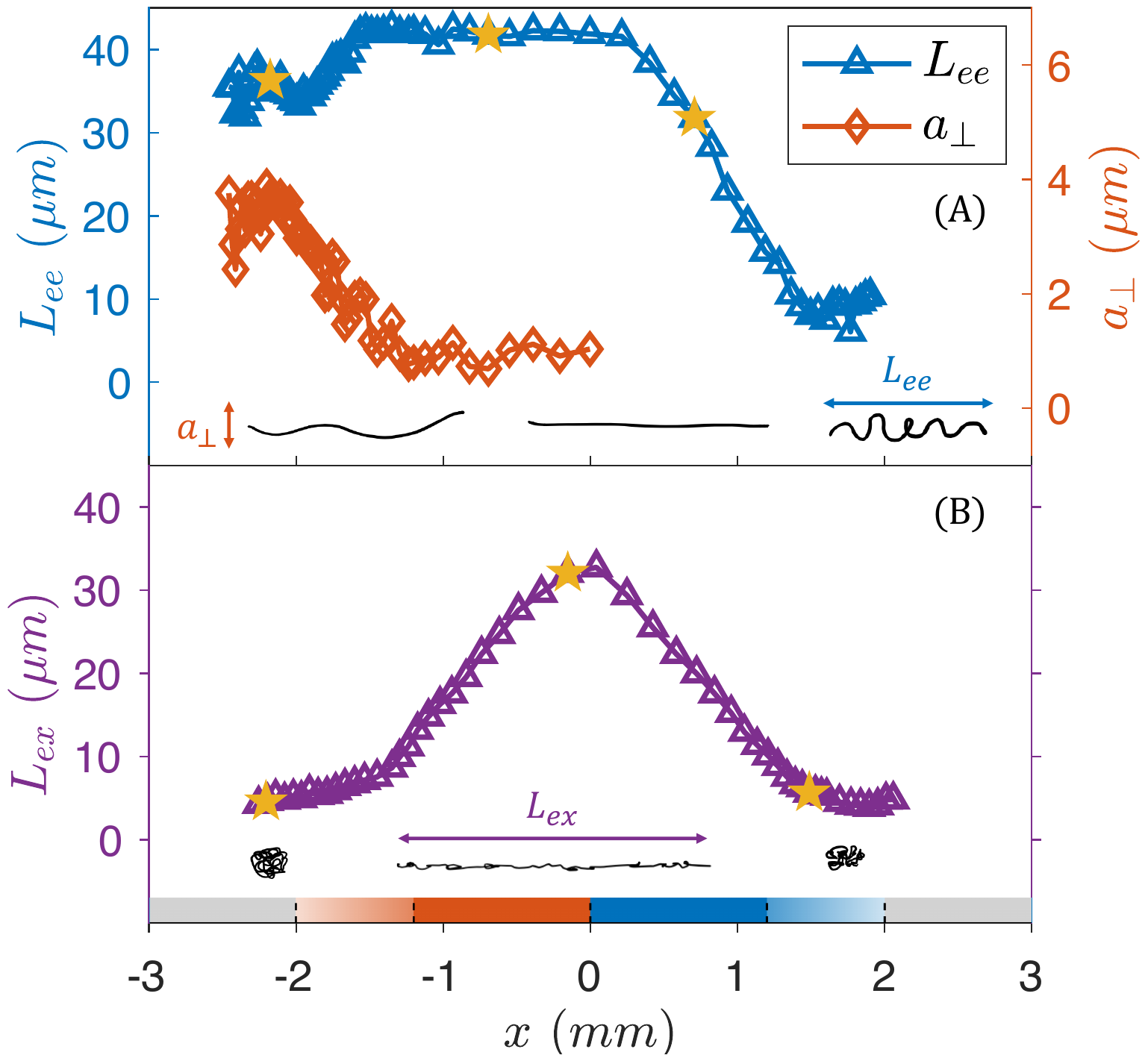}
	\end{center}
	\caption{Quantitative measurements of the dynamics of individual actin filament and DNA molecule with maximum strain rate $\dot{\epsilon}=0.6 s^{-1}$. (A): The end-to-end distance $L_{ee}$ and shape fluctuation amplitude $a_{\perp}$ of an actin filament as a function of position, with a contour length of $42\mu$m. (B): Molecule extension $L_{ex}$ of a DNA molecule as a function of position, with a maximum extension of $32\mu$m. Schematic drawings of polymer conformations in different parts of the channel, their positions are marked with yellow stars.}
	\label{fig_endToend}
\end{figure}

\begin{figure}[!b]
	\centering
	\includegraphics[width=0.99\columnwidth]{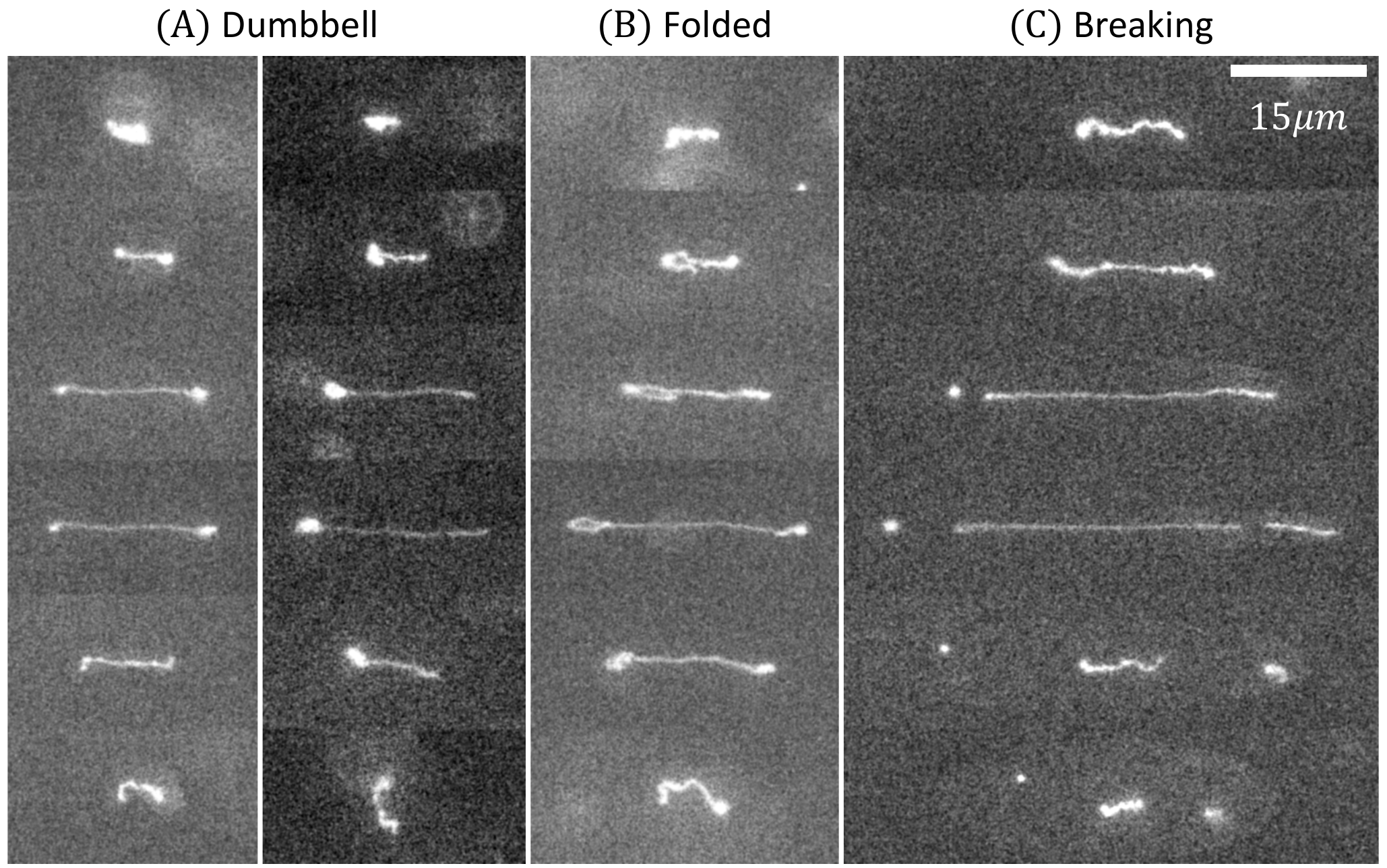}
	\caption{Different dynamics of DNA molecules during extension (four top rows) and compression (three bottom rows). (A): double end and single end dumbbell shape; (B) folded shape; (C) breaking into three fragments during extension.}
	\label{fig_DNA_dynamics}
\end{figure}

The image quality allows for quantitative characterisation of the different polymer morphologies. \Cref{fig_endToend}\figcol{A} shows measurements of the dynamics of an actin filament. The blue triangles represent the end to end distance $L_{ee}$ which reflects the overall filament shape. $L_{ee}$  increases slightly when the filament enters the extension flow as Brownian shape fluctuations are suppressed by the viscous flow and the filament is straightened out. Consequently, at the same time, the transverse amplitude of shape fluctuations $a_{\perp}$ (represented by the red diamonds) decreases. Both $L_{ee}$ and $a_{\perp}$ quickly plateau reflecting the full suppression of bending fluctuations and the subsequent inextensibility of the backbone. 
A fundamentally different behaviour is now observed when the straight filament enters the compressive part where it undergoes a buckling instability~\cite{chakrabarti2020flexible}. This is characterised by a sharp decrease of $L_{ee}$ while a 3D-coiled structure is formed and more and more compressed by the flow. This compact state will relax back to the initial straight, fluctuating conformation in the downstream part of the channel.

\Cref{fig_endToend}\figcol{B} presents the dynamics of a DNA molecule. When entering the extension part of the flow a strong increase of the molecule extension $L_{ex}$ is observed, reflecting the fact that the molecule transitions from a coiled state to a stretched conformation. The increase of the molecule extension $L_{ex}$ along the channel (and thus over time) reveals the stretching dynamics in extension flow. A change in these dynamics is clearly visible when the polymer enters the region of constant and maximum strain rate, reflecting the change in extension properties. Once the DNA molecule reaches the compressive part it is pushed back into its initial conformation. Note that the initial molecule extension measured is larger than the radius of gyration of the DNA molecule at rest. This is most likely due to the history of the DNA molecule in the upstream channel, either the initial stretching of the molecule in the flow focusing part of our device did not completely relax back to the equilibrium position or it experienced small shear in the straight part of the channel if not perfectly centred.

The very high image quality obtained through the tracking technique gives access to the quantitative details of the dynamics but also to the different morphologies obtained. As an example, different stretching dynamics can clearly be observed for DNA molecules in \cref{fig_DNA_dynamics}. One can distinguish molecule stretching with one or two coiled ends, while folded dynamics reveal the variety of dynamics already observed in \cite{perkins1997single,smith1998response} and reminiscent of the concept of molecular individualism pointed out by de Gennes~\cite{de1997molecular}. We observe in a few cases breakage during extension that might be facilitated by photodamage. Note also that the advantage of our set-up is that in all cases it provides precise information on the exact conformation of the polymer when entering the straining part of the channel, allowing thus for a precise investigation of stretching dynamics as a function of flow history or initial state.

\begin{figure}[!t]
	\begin{center}
		\includegraphics[width=0.7\columnwidth]{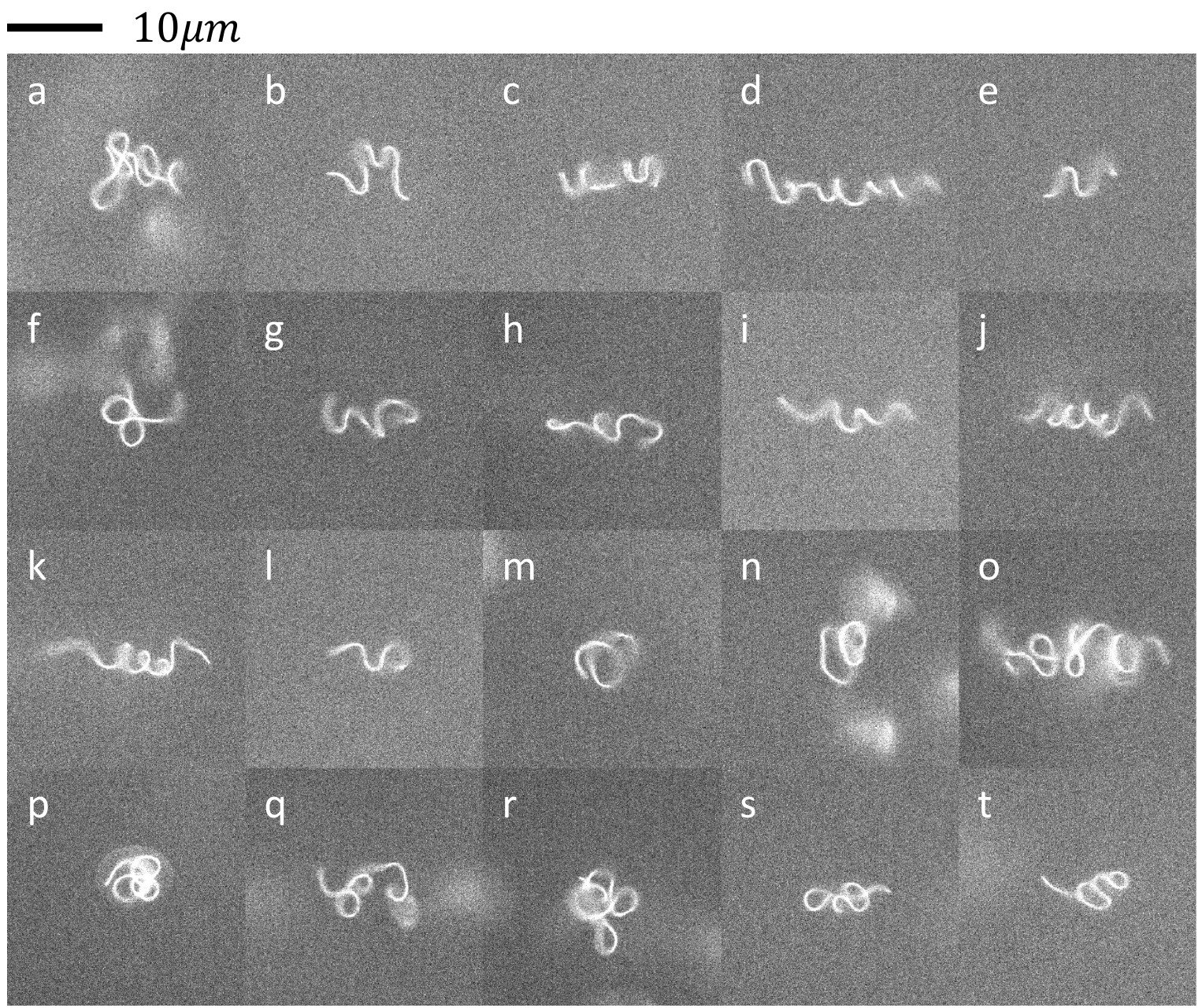}
	\end{center}
	\caption{Various shapes and structures of different actin filaments are observed in the compression flow. They include helicoidal shapes as seen on snapshots (d), (e), (k) and (n) or more complex shapes probably including knots as seen on snapshots  (a), (p) and (r).}
	\label{fig_coiled}
\end{figure}

In a similar way, \cref{fig_coiled} shows multiple conformations of actin filaments towards the end of the compressive flow part. One recognises regular helices, more coiled conformations and even knots.

\subsection{Protein aggregates}

\begin{figure}[!t]
	\centering
	\includegraphics[width=\columnwidth]{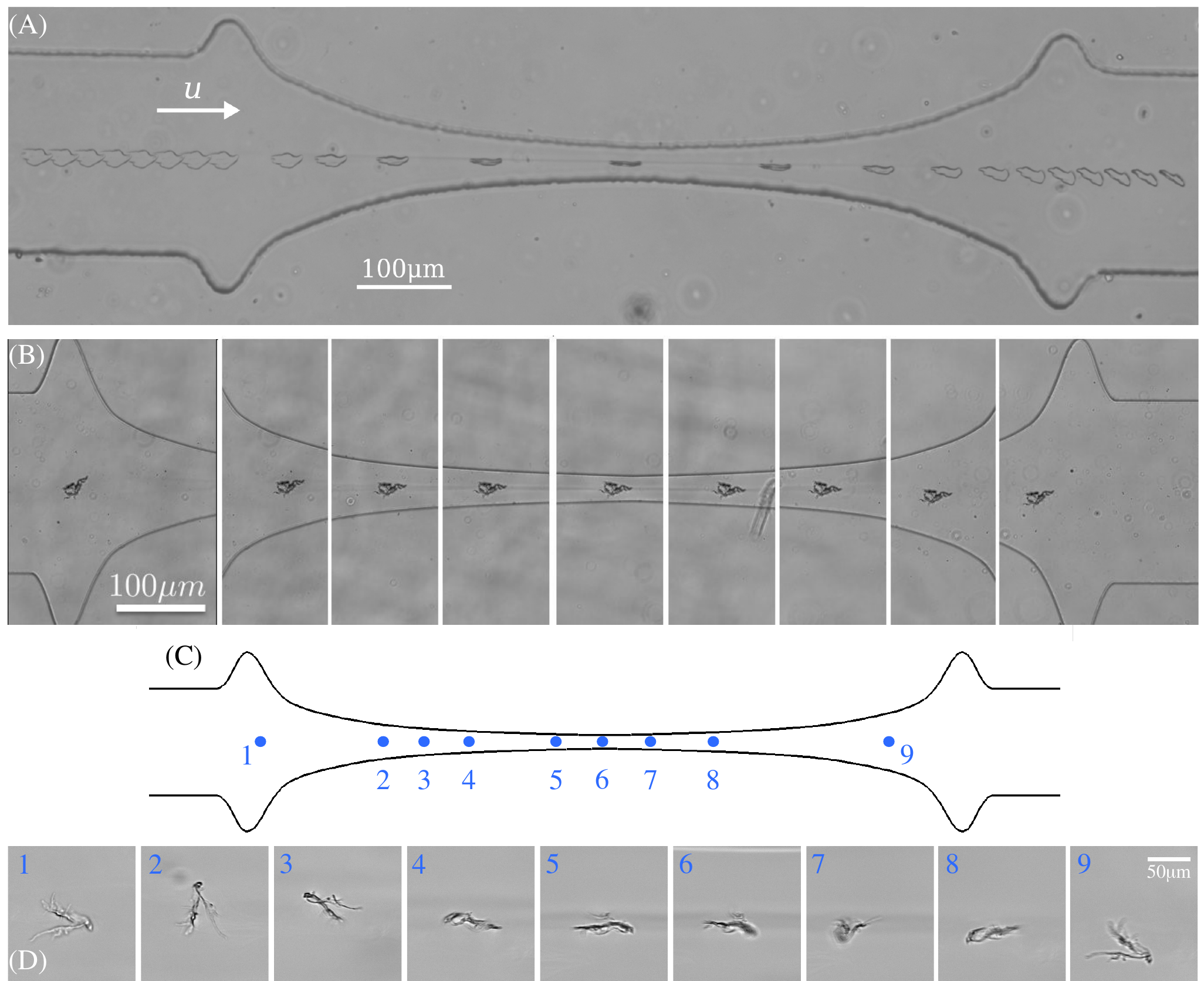}
	\caption{(A) Protein aggregate transported in a channel of AR=2 and nominal dimensions $H=100\mu$m, $w_u=200 \mu$m, $w_c=25 \mu$m. Phase contract microscopy is used, with a small magnification objective (x10) and a fixed stage. 
(B) Protein aggregate transported in a channel of AR=8 and nominal dimensions  $H=100\mu m$,  $w_c=100\mu m$ and $w_u=800\mu m$. Phase contract microscopy and a small magnification objective (x10) are used together with the tracking method.
(C\&D) Protein aggregate transported in a channel of AR=8 and nominal dimensions  $H=100\mu m$,  $w_c=100\mu m$ and $w_u=800\mu m$. Phase contrast microscopy, a x40 lens and the tracking method are used.  The positions in the channel and snapshots of aggregates are illustrated schematically respectively in (C) and (D).}
	\label{Fig:aggregates0}
\end{figure}

Protein aggregates occur in a large variety of shapes and sizes depending on their fabrication conditions \cite{Engelsman2011}. Little is known on their dynamics under flow, despite this being crucial information to understand clogging of syringes, the functioning of auto-injectors or filling processes. The excellent observation capabilities of our set-up can lead to such detailed information under transport in well controlled straining flows (\cref{Fig:aggregates0}). The aggregates used here are estimated to have an elastic modulus of 0.1MPa \cite{Duchene2020}. Typical stresses exerted on transported particles in our flow conditions are ${\dot \epsilon} \eta \approx 0.1 $Pa, several orders of magnitude smaller than typical moduli and as a consequence no significant aggregate deformation is observed. However, due to their larger and more compact shape compared to the biopolymers described previously they explore the three dimensional nature of the flow, leading to very rich reorientation dynamics.

In \cref{Fig:aggregates0}\figcol{A} a relatively compact aggregate is shown while being transported through an optimised converging/diverging channel. Given their size, protein aggregates can be observed using smaller magnification objectives if necessary, which allows capturing a larger portion of the channel within the field of view. In addition, protein aggregates can be observed using phase contrast together with short exposure times (around 1ms) limiting image blur. The second channel geometry, with AR=2 ($H=100\mu$m, $w_u=200 \mu$m, $w_c=25 \mu$m), is used here and observations with a fixed stage have been performed still yielding reasonable image quality. The smaller magnification in combination with the reduced channel size permits to see the channel walls on the images and thus to visually locate the aggregate position during transport.  This geometry has the particularity of changing aspect ratio from above 1 to below 1 when going from the straight channel towards the throat. This means that in the straight part of the channel, the largest dimension of the rectangular cross-section is the width, while in the throat the height is largest. The transported aggregate adapts to this change in channel aspect ratio and rotates sideways when passing the throat. It can now be seen as a narrow line from our observation perspective, revealing its disc like shape. Downstream of the throat the aggregate rotates back into the $xOy$ plane reaching an orientation close to the one observed upstream of the throat.

In \cref{Fig:aggregates0}\figcol{B} an aggregate of complex shape is shown, now transported in the channel or AR=8 still using a small magnification. For improved image quality the tracking method described in Section~\nameref{sec_implementation} is used here and at each instant the details of the aggregate are clearly visible, revealing that the particle becomes more and more aligned with the direction of extension while transported towards the throat. Downstream of the throat the aggregate slowly rotates back and looses the full alignment.  

Even more complex shaped aggregates can lead to even more complex reorientation dynamics. An illustration of such effects is given in \cref{Fig:aggregates0}\figcol{D}, showing the transport of an extended complex-shaped aggregate that resembles a filigree. This figure illustrates again the efficiency of our tracking system to provide sharp images where all details of the particle can be identified and followed, now using a higher magnification objective (x40) and the channel of AR=8. Tumbling in the  $xOy$ plane, probably induced by the fact that the aggregate is not perfectly on the centreline, alignment with the direction of extension as well as rotation out of the $xOy$ plane can be observed as before. 

The interplay of specific and potentially complex aggregate shapes and the three dimensional flow geometry lead to various reorientation dynamics highlighted by the examples discussed. Due to the fact that most injection devices contain constrictions the investigation of the flow dynamics in our device can lead very relevant information on the transport of aggregates in such devices and as a consequence the possible obstruction of needles or syringes.

\section{Discussion}
\label{Disc}

In this section we discuss the typical operation range of our set-up in terms of extension rates and applied stresses, as well as its limitations and possible improvements.

An important parameter in our experiments is the total Hencky strain, as defined in Equation~\cref{eq_profile_SR}. The Hencky strain is calculated as the integral of the effective strain rate over time, and characterises the accumulated strain of a macromolecule traveling along the centreline of the converging region and is a function of the channel geometry, but independent of the kinematics. The total Hencky strain is therefore fixed for each geometry and needs to be defined at the optimisation stage. The strain rate at the centreline on the other hand ${\dot{\epsilon}}\approx \frac{u_c-u_u}{l_h+0.5l_v}$, is a function not only of the geometrical parameters of the designed channel, namely the contraction ratio and the length of the straining parts (and to a less extent the AR in case of 3D planar geometries), as well as a linear function of the applied flow rate. For 2D channels the strain rate can be approximated as ${\dot{\epsilon}}\approx \frac{u_c}{l_h+0.5l_v}(1-\frac{1}{CR})$, here expressed in terms of the maximum velocity obtained in the throat. Our tracking method currently limits the maximum velocity in the throat to 1mm$/$s, leading to an upper limit in strain rate of around ${\dot{\epsilon}}\approx 5s^{-1}$ for the geometries used here. This limitation stems from the technical specifications of our tracking method and namely the frequency with which signals are sent to the stage to control its movement and can certainly be improved. Note that one should also make ensure the flows remains in the small Reynolds number limit when increasing the flow rates. Alternatively, the channel geometry can be modified to increase the strain rate, by for example increasing CR or reducing the length of the straining parts of the channel. In practice, large CRs lead to very wide channels upstream of the straining parts, limiting again possible increases of the strain rate. This is particularly true when maintaining the throat rather wide to avoid influence of the side walls on the particle transport dynamics. We estimate possible increases in strain rate to about a factor of ten while remaining in the framework of our optimised hyperbolic channels and tracking methodology. 

Typical strain rates achieved in our device are perfectly suited to study the deformation of long polymers as shown here for DNA or actin filaments. These molecules deform when the time scale given by the inverse of the strain rate becomes comparable to their relaxation times, typically of the order of seconds. Note that increasing the viscosity of the suspending fluid increases the relaxation time, making the effects of deformation visible at smaller strain rates.

When considering more compact objects, as for example the protein aggregates presented here, the ratio between the applied viscous stresses and their elastic modulus determines the importance of their deformation. The relatively small strain rates achieved in our device also lead to small viscous stresses, the latter being proportional to the strain rate and the viscosity $\sigma={\dot{\epsilon}} \eta$. Typical viscous stresses achieved in our study are around 10-100 mPa. While for inert particles viscosities of up to a hundred times the viscosity of water can be used, this might not be a tunable parameter for bio-particles that often require very specific buffer solutions, with typical viscosities close to the viscosity of water. Compact objects can consequently only be significantly deformed for very low moduli, where a 1$\%$ deformation can be reached for moduli of 1-10Pa, in agreement with the fact that we could not detect aggregate deformation in our study. 

Another very important geometrical parameter that can be tuned at will is the length of the transition region. This region is essential to avoid overshots in strain rate that could damage fragile molecules, but its length can be modulated. A very short transition region will avoid molecule pre-stretch, whereas a longer region will lead to a well defined and controlled pre-stretch.

In this work, we made use of two specific optimised designs with particular geometrical characteristics ($CR$, $AR$, ${l}_{h}$, ${l}_{v}$), but these can be easily customised according to the application and the bio-particle under study. Using the described optimisation procedure we can obtain tailored shapes that allow, within certain limits, for specific strain rates and Hencky strain relevant for the application under consideration.

\section{Conclusion}
\label{Conc}

In this work, we demonstrated the performance of optimised microfluidic geometries for the mechanical characterisation of bio-particles under well-controlled straining flow. The specific geometry of the microfluidic channel was obtained from a numerical optimisation process and is able to generate a constant strain rate along the flow centreline over extended distances, in contrast with other commonly employed configurations. The flow kinematics have been validated experimentally and a sophisticated tracking method was developed for displacing the microscope stage in such a way that the bio-particle under observation is always in focus and within the image frame. High image quality is obtained in this way, even for single molecules under transport with cross scale varying velocities. 

Besides the obvious advantage of obtaining extended regions of homogeneous straining flow, one of the key benefits of our set-up is that it provides precise information on the size/shape/conformation of the particles at the entrance of the straining region, allowing for the investigation of effects of flow history or initial state on the deformation/stretching/orientation dynamics under homogeneous straining flow. Furthermore, the combination of a converging and a diverging part leads to an extension flow that is followed by a compression flow in the same device, making it particularly versatile.

The microfluidic device we have developed allows the observation of bio-particles of interest in biological systems that exhibit very different conformations and properties. First, two different bio-polymers, DNA molecules and actin filaments were used. Their conformations at rest are very different, a coiled state for the first and an extended state for the latter. The unique experimental design used here enabled us to confront in this way two fundamental morphological transitions of polymers:  the buckling instability of straight polymers induced by compressive viscous forces and the well-known coil-stretch transition induced by extensional viscous forces on coiled polymers. Protein aggregates on the other hand did not deform under the viscous forces applied here, but showed very rich reorientation dynamics.

In summary, the optimised microchannel in combination with the tracking algorithm offers the possibility to investigate the qualitative and quantitative behaviour of numerous micronscale bio-particles of different properties and to obtain insight into their mechanical properties and dynamics under flow relevant in a wide range of applications.

\section*{Conflicts of interest}
There are no conflicts to declare.
 
\section*{Acknowledgements}
AL, OdR, YL , JF and CD acknowledge funding from the ERC Consolidator Grant PaDyFlow (Agreement No. 682367). YL acknowledge funding from Natural Science Basic Research Plan in Shaanxi Province of China (No. 2020JQ-566).


\balance

\renewcommand\refname{References}
\bibliography{rcs} 
\bibliographystyle{rsc} 

\end{document}